\documentclass[a4paper, 12pt]{amsart}
\usepackage{amsmath, amssymb, amsthm}
\usepackage[english]{babel}
\usepackage[T1]{fontenc}
\usepackage{graphicx}
\usepackage{rotating}
\usepackage{array,multirow}
\usepackage{url}
\newcommand{\R}{\mathbb R}
\newcommand{\N}{\mathbb N}

\def\be#1\ee{\begin{equation}#1\end{equation}}
\newcommand{\fer}[1]{(\ref{#1})}

\numberwithin{equation}{section}

\newcommand{\bq}{\begin{equation}}
\newcommand{\eq}{\end{equation}}
\newenvironment{equations}{\equation\aligned}{\endaligned\endequation}
\def\bqa{\begin{eqnarray}}
\def\eqa{\end{eqnarray}}

\def\e{\epsilon}

\def\t{\tau}

\def\bx{{\bf x}}

\def\bm{{\bf m}}
\def\bxi{{\pmb\xi}}

\def\aa{{\bf a}}
\def\ab{{\bf b}}

\newcommand{\bd}{\begin{displaymath}}
\newcommand{\ed}{\end{displaymath}}
\newcommand{\ba}{\begin{eqnarray}}
\newcommand{\ea}{\end{eqnarray}}

\def\ff{\widehat f}

\def\gg{\widehat g}
\def\hh{\widehat h}

\def\N{\mathbb{N}}
\def\R{\mathbb{R}}

\theoremstyle{plain}

\title{Measuring multidimensional heterogeneity in emergent social phenomena}
\author{Giuseppe Toscani}

\thanks{Department of Mathematics  of the University of Pavia, and IMATI CNR, Pavia, Italy.  \\ e.mail: {\tt giuseppe.toscani@unipv.it}. 
}

\date{\today}

\begin{document}
\maketitle

\begin{center}\small
\parbox{0.85\textwidth}{

\textbf{Abstract.} 
 Measuring inequalities in a multidimensional framework is a challenging problem which is common to most field of science and engineering. Nevertheless, despite the enormous amount of researches illustrating the fields of application of inequality indices, and of the Gini index in particular, very few consider the case of a multidimensional variable. In this paper, we consider in some details a new inequality index, based on the Fourier transform, that can be fruitfully applied to measure the degree of inhomogeneity of multivariate probability distributions. This index  exhibits a number of interesting properties that make it very promising in  quantifying the degree of inequality in data sets of complex and multifaceted social phenomena. 


\medskip

}
\end{center}

\section{Introduction}

Among other approaches, the description of  social phenomena in a multi-agent system can be successfully obtained by resorting to statistical physics, and,  in particular, to methods borrowed from kinetic theory of rarefied gases. The main goal of the mathematical modeling is to construct master equations of Boltzmann type, usually referred to as kinetic equations, suitable to describe the time-evolution of some \emph{social} characteristic of the agents, like wealth, opinion, knowledge, or others  \cite{CCCC,NPT,PT13}.

The building block of kinetic theory is represented by the details of microscopic interactions, which, similarly to binary interactions between particles velocities in the classical kinetic theory of rarefied gases, describe the elementary variation law of the selected agent's traits.  Then, the kinetic description consequent to the microscopic law of variation  is able to capture both the time evolution  of the number density and the steady profile, an important equilibrium distribution that should resume at best the characteristics of the phenomenon under investigation. 

Once the emergent steady profile relative to a social phenomenon has been identified, various features allow to have a more precise measurement of its social characteristics,  to better understand in this way the macroscopic effect of the microscopic behavioral interactions of agents.

Among the various features that can be introduced to measure properties of equilibria emerging from kinetic equations modeling social phenomena, a relevant importance has been assumed by inequality indices, quantitative scores that take values in the unit interval, with the zero score characterizing perfect equality.

To better clarify the point, we refer to a classical example provided by the kinetic description of wealth distribution in a western society.
Among the kinetic models introduced in recent years to study the evolution of wealth distribution in a multi-agent society \cite{PT13}, a Fokker--Planck type equation assumed a leading role. This equation, that reads
  \be\label{FP}
 \frac{\partial f}{\partial t} = \frac \sigma{2}\frac{\partial^2 }
 {\partial w^2}\left( w^2 f\right) + \lambda \frac{\partial }{\partial w}\left(
 (w-1) f\right),
 \ee
 describes the evolution of the wealth distribution density $f(w,t)$  towards a steady state.\\
In \fer{FP}  $\lambda$  and $\sigma$  denote two positive constants related to essential properties of the trade rules of the agents, linked to the saving propensity and, respectively, the risk.  Equation \fer{FP} has been first derived  by Bouchaud and Mezard \cite{BM} through a mean field limit procedure applied to a stochastic dynamical equation for the wealth density.  The same equation was subsequently obtained by  the present authors with Cordier and Pareschi \cite{CoPaTo05} via an asymptotic procedure from a Boltzmann-type kinetic model for trading agents. \\
 The unique stationary solution of unit mass of \fer{FP} is given by the inverse
 Gamma distribution \cite{BM,CoPaTo05}
 \be\label{equi2}
f_\infty(w) =\frac{(\mu-1)^\mu}{\Gamma(\mu)}\frac{\exp\left(-\frac{\mu-1}{w}\right)}{w^{1+\mu}},
 \ee
  where
  $$ \mu = 1 + 2 \frac{\lambda}{\sigma} >1.
$$
This stationary distribution,  as predicted by the  analysis of the italian economist Vilfredo Pareto \cite{Par}, exhibits a power-law tail for large values of the wealth  variable.\\
In this context, the classical feature is to quantify the degree of economic \emph{inequality} contained in the wealth distribution associated to this equilibrium shape in terms of the parameters $\lambda$ and $\sigma$, a quantification that is usually done by resorting to  the Gini index, a well-known measure of inequality first proposed by the Italian statistician Corrado Gini more than a century ago \cite{Gini1,Gini2}. \\
In economics,  inequality indices quantify the socio-economic divergence of a given wealth measures from the state of perfect equality. Their relevance is certified by the fact that, in addition to Gini index,   many other inequality indices have been proposed to classify wealth measures \cite{BL,Cou,Cow,HN}. 

However, as recently discussed in \cite{Ban,Eli,Eli3}, the challenge of measuring the statistical heterogeneity of  measures is not limited to economics, but arises in most fields of science and engineering, and it is one of the fundamental features of data analysis.  

A marked limitation in this type of analysis is that the inequality indices mainly used in the literature work well for one-dimensional features, while their extension to many dimensions presents several difficulties.  This is in contrast with the fact that in several problems arising from socio-economic phenomena the prevailing interest is related to understanding multidimensional phenomena. \\ Remaining in the field of the kinetic description of socio-economic phenomena, we quote here some examples in which the social aspects of the society under study are intimately connected, and have been treated resorting to a kinetic framework that naturally give rise to multivariate equilibria. The first one refers a kinetic equation for the evolution of the probability distribution of two goods among a huge population of agents \cite{TBD}, where binary exchanges are characterized by Cobb-Douglas utility functions  and the Edgeworth box for the description of the common exchange area in which utility is increasing for both agents. This leads to  a drift-diffusion equation of Fokker-Planck type in two dimensions for the joint distribution of the two goods. \\ The second example is related to a deep understanding of the joint action of knowledge and wealth in the formation of stationary wealth profiles \cite{PT14}. There, the underlying Fokker--Planck equation drives the system towards a steady profile which depends of both the knowledge and wealth variables. \\ The last example refers to a fully coupled mathematical model in which knowledge and social status of individuals in a western society influence each other \cite{DTZ}. Also in this case, one has to deal with a bivariate equilibrium profile from which one would extract global informations without resorting to one-dimensional inequality measures applied to the marginal distributions.
 
 Measuring inequalities in a multidimensional framework is a question which is nowadays a priority also in the European agenda \cite{Eu, Eu2, Kov}.  Indeed, as outlined in introduction to this action, "the pursuit of a more equal and fairer Europe requires extensive knowledge on prevailing inequalities across multiple life domains. Inequality is a complex and multifaceted phenomenon, and every attempt to assess multidimensional inequalities comes with a number of conceptual and empirical challenges. For example, inequality and poverty do not necessarily move in the same direction: low poverty levels in a society may be combined with high inequality due to large differences between those at the top and those in the middle of the distribution. Against this backdrop, the EU Multidimensional Inequality Monitoring Framework aims to contribute to the measurement, monitoring and analysis of a wide range of different aspects of inequality".\footnote{https://composite-indicators.jrc.ec.europa.eu/multidimensional-inequality}

 Currently, the literature on multidimensional inequality measures is vast, as can be seen by taking a look at the extensive references of some recent contributions \cite{AA,AZ}.  However, as listed below, most of these approaches to multidimensional indices are based on classical arguments,  derived by classical economical indices. 

 An indispensable tool to build inequality indices  is the Lorenz function and its graphical representation, the Lorenz curve \cite{Lor}. The Lorenz curve plots the percentage of total income earned by the various sectors of the  population, ordered by the increasing size of their incomes.
The Lorenz curve is typically represented as a curve in the unit square of opposite vertices in the origin of the axes and the point $(1,1)$,  starting from the origin and ending at the point  $(1,1)$. 
  
The diagonal of the square exiting the origin is the line of perfect equality, representing a situation in which all individuals have the same income. Since the diagonal is the line of perfect equality, the closer the Lorenz curve is to the diagonal, the more equal is the
 distribution of income. 
 
 This idea of \emph{closeness} between the line of perfect equality and the Lorenz curve can be expressed in many ways, each of which gives rise to a possible measure of  inequality.
  Thus, starting from the Lorenz curve, several indices of  inequality can be defined, including the Gini index \cite{Gini1,Gini2}. Various indices were obtained by looking at the maximal distance between the line of perfect equality and the Lorenz curve, either horizontally or vertically, or alternatively parallel to the other diagonal of the unit square \cite{CGC, Eli}.

Starting from this framework, different multivariate indices have been proposed in the pertinent literature.   Most of them are based on the notion of Lorenz zonoid \cite{KM} a multi-dimensional generalization of the Lorenz curve.

Despite the enormous amount of research illustrating the fields of application of inequality indices, the use of arguments based on Fourier transforms appears rather limited. In particular, although the Gini index can be easily expressed in terms of the Fourier transform,  its expression in Fourier seems not  considered at all in applications.

The importance of expressing inequality measures in terms of the Fourier transform of measures has been recently outlined  in \cite{To1},  not only by expressing well-known one-dimensional inequality measures in terms of the Fourier transform, but introducing and studying a novel inequality index directly expressed in terms of the Fourier transform. 

In the rest of the paper, we will show how this new inequality measure can be easily generalized to cover multivariate probability distributions, by enlightening its main properties.  We restrict the forthcoming analysis to theoretical considerations only,  referring the interested reader to its application in the field of multivariate statistics \cite{GRT}.

\section{A new inequality index for multivariate distributions} \label{sec:nd}

The goal of this Section is to present in some details a novel inequality index which can be applied to measure the heterogeneity of multivariate distributions \cite{GRT}. This index is obtained by suitably generalizing a new one-dimensional index introduced in \cite{To1}. 

In what follows, for a given $1\le n \in \N$, we denote by $P_s(\R^n)$, $s \ge 1$, the class of all probability measures $F = F(\bx)$ on the Borel subsets of $\R^n$  such that
\[
m_s(F)  = \int_{\R} |\bx|^s dF(\bx) < + \infty,
\]
where, for a given column vector ${\bf v}$ of dimension $n$,  ${\bf v}^T = (x_1,x_2,\dots,x_n)= \bx $ is a point in $\R^n$, and $|{\bf v}|= |\bx| =\sqrt{\bx{\bf v}}$ is the modulus of the vector, i.e. the distance of the point $\bx$ from the origin of the cartesian axes. \\ Further, we denote by $ \tilde P_s (\R^n)$ the class of probability measures $F \in P_s(\R^n)$ which possess a mean value vector $\bf m$ with positive components $m_k$, $k =1,2,\dots, n$, i.e.
\[
{m}_k (F) =  \int_{\R^n } x_k \, dF(\bx) >0, \quad k =1,2,\dots, n,
\]
and with $P_s^+(\R^n)$ the subset of probability measures $F \in P_s(\R^n)$ such that $F(\bx) = 0$ if at least one component $x_k \le 0, k =1,2,\dots, n$. 

On the set  $\tilde P_s(\R^n)$ of probability measures , we consider the set $\mathcal F_s^n $  of their $n$-dimensional Fourier transforms, where, for $F=F(\bx) \in  \tilde P_s(\R^n)$,
\be\label{fou}
\ff(\bxi) = \int_{\R^n}  e^{-i \bx\bxi} \, dF(\bx).
\ee
In \fer{fou} we denoted by $\bxi$ the $n$-dimensional column vector of components $\xi_k, k =1,2,\dots, n$. 

When $n =1$,  in alternative to well-known inequality indices,  for a given distribution $F \in P_s(\R)$, the following measure of heterogeneity was proposed in \cite{To1}
\be\label{ine-T} 
T(F) = \frac 1{2m} \sup_{\xi \in \R} \left|  \left. \frac{d\ff(\xi)}{d\xi}\right|_{\xi =0} \ff(\xi) - \frac{d\ff(\xi)}{d\xi} \right|. 
\ee
 In definition \fer{ine-T},  $m>0$  denotes the mean value of the distribution $F$. \\
 Apparently, the index \fer{ine-T} is completely disconnected from others most used indices, including the well-known Gini index \cite{Gini1,Gini2}, strongly related to Lorenz curve \cite{Eli}.  However, looking at Gini index from ther Fourier transform side,  an interesting relationship appears. \\
Let us consider a probability measure  $F \in P_s^+(\R)$, of mean value $m>0$.  As shown in \cite{To1}, the classical Gini index 
\be\label{Gin}
G(F) = 1 - \frac 1m\int_{\R_+} (1- F(x))^2\, dx.
\ee
can be expressed in terms of one-dimensional Fourier transform as follows:
\be\label{Gin-F}
G(F) = 1 - \frac 1{2\pi m} \int_\R \frac{|1 -\ff(\xi)|^2}{|\xi|^2} \, d\xi.
\ee
Expression \fer{Gin-F} clarifies that the Fourier expression of the classical Gini index is a function of a certain distance between probability measures $F$ and $G$ \cite{ToTo}, namely
\[
d_2(F,G) = \int_\R \frac{|\ff(\xi)-\gg(\xi)|^2}{|\xi|^2} \, d\xi.
\]
Resorting to this analogy, in \cite{To1} new inequality measures have been introduced, some of them related to the supremum distance
\be\label{d-inf}
d_\infty(F,G) = \sup_{\xi\in \R} \frac{|\ff(\xi)-\gg(\xi)|}{|\xi|}.
\ee
This type of metrics have been extensively studied in connection with the convergence to equilibrium of kinetic equations, as  alternatives to more classical entropies \cite{GTW,CaTo,ToTo}. \\
Let $X$ be a random variable, of mean value $m>0$ characterized by a differentiable probability measure $F \in P_s^+(\R)$, and denote by $f(x) = dF(x)/dx$ its probability density. In this case, in addition to the classic expression \fer{Gin},  Gini index can be expressed in alternative forms, one of which is particularly interesting to enlighten contact points with the $T$-index defined by \fer{ine-T}.  This alternative expression reads
\be\label{Gin-2}
G(F) = 2 \int_{\R_+} (1- F(x))\left( f(x) -\frac xmf(x)\right)\, dx.
\ee
Indeed,
\[
\int_{\R_+} (1- F(x)f(x) \, dx = - \frac 12 \int_{\R_+} \frac d{dx}(1-F(x)^2\, dx =  - \frac 12 \left.(1-F(x)^2\right|_0^\infty =\frac 12,
\]
while, integrating by parts
\[
\int_{\R_+} (1- F(x)f(x)\frac xm \, dx = - \frac 12 \int_{\R_+}\frac xm\, \frac d{dx}(1-F(x)^2\, dx = \frac 1{2m} \int_{\R_+}(1-F(x)^2\, dx.
\]
Consequently, if we set $H(x) = 1-F(x)$, thanks to Plancherel identity we obtain
\begin{equations}\label{Gin-F2}
G(F) = & 2 \int_{\R_+} (1- F(x))\left( f(x) -\frac xmf(x)\right)\, dx = \\
& \frac 1{m\pi}\int_\R \overline{\widehat H(\xi)} \left( \left. \frac{d\ff(\xi)}{d\xi}\right|_{\xi =0} \ff(\xi) - \frac{d\ff(\xi)}{d\xi} \right),
\end{equations}
where $\overline{\widehat H(\xi)}$ is the complex coniugate of the Fourier transform of $1-F(x)$, given by
\[
H(\xi) = \frac{1 -\ff(\xi)}{i\xi}.
\]
Consequently, the value of the Gini index depends on the product  of two different quantities, working  in opposite directions. Indeed, according to \fer{d-inf}, the term  $\overline{\widehat H(\xi)}$ quantifies the distance of $\ff(\xi)$ from the state of perfect inequality, represented by the value $1$, the Fourier transform of a Dirac delta function located in $x=0$. On the contrary, the term
\[
\left. \frac{d\ff(\xi)}{d\xi}\right|_{\xi =0} \ff(\xi) - \frac{d\ff(\xi)}{d\xi},
\]
 that vanishes in correspondence to $e^{-im\xi}$, the Fourier transform of a Dirac delta function localized in the mean value $m$ of $f(x)$, quantifies the distance of $\ff(\xi)$ from the state of perfect equality. This clarifies both the nonlinearity of Gini index, and the advantages of the choice of the inequality index \fer{ine-T} as alternative measure of the heterogeneity of the distribution. A further advantage is represented by the possibility to easily extend the measure \fer{ine-T} to higher dimensions.

Following  \cite{GRT},  we introduce on $\mathcal F_s^n$ the multivariate inequality index $T_n(F)$, expressed by the formula
\be\label{ine-Tn} 
T_n(F) = \frac 1{2|\bm|} \sup_{\bxi \in \R^n} \left| \nabla \ff(\bxi = {\bf 0}) \ff(\bxi) - \nabla \ff(\bxi) \right|.
\ee
 In definition \fer{ine-Tn},  $\nabla \ff(\bxi)$ denotes the gradient  of the scalar function $\ff(\bxi)$. Indeed,  $F \in P_s(\R^n)$ implies that  $\ff(\bxi)$ is  continuously differentiable. \\

It is immediate to show that the functional $T_n(F)$ is invariant with respect to the scaling (dilation)
\[
F(\bx) \to F(c\bx), \quad c >0.
\]
moreover, as shown in \cite{To1} for the one-dimensional index,  $T_n$ is bounded from above by $1$. 
Indeed, since for any given $F \in P_s^+(\R^n)$ it holds $|\ff(\bxi)| \le \ff({\bf 0}) = 1$, and 
\[
\frac{\partial \ff(\bxi)}{\partial \xi_k} = -i \int_{(\R_+)^n}  x_k e^{-i \bx\bxi} \, dF(\bx), \quad k =1,2,\dots, n,
\]
one obtains the bound
\be\label{posi}
\left| \frac{\partial \ff(\bxi)}{\partial \xi_k}\right| \le \int_{(\R_+)^n}  x_k \left|e^{-i \bx\bxi}\right| \, dF(\bx) =m_k,
\ee
which implies $|\nabla \ff(\bxi)| \le |\nabla \ff(\bxi ={\bf 0})| = |\bm|$.

Hence,  by the triangular inequality one concludes that  $T_n(F)$ satisfies the usual bounds
\be\label{bound-n}
0 \le T_n(F) \le 1,
\ee
and $T_n(F) = 0$ if and only if $\ff(\bxi)$ satisfies the differential equations
\[
\frac{\partial \ff(\bxi)}{\partial \xi_k} = \left. \frac{\partial \ff(\bxi)}{\partial \xi_k}\right|_{\bxi = \bf 0} \ff(\bxi), \quad k =1,2,\dots, n,
\]
with $\ff({\bf 0}) = 1$. 

Thus, as in the one-dimensional case,   $T_n(F)$ vanishes  if and only if $\ff(\bxi) = e^{-i\bm\bxi}$, namely if $\ff(\bxi)$ is the Fourier transform of a Dirac delta function located in the point $\bx =\bm^T(F)$. Note however that, even if the functional $F$ is defined in the whole class $\tilde P_s(\R^n)$,  the upper bound is lost if the probability measure $F \notin  P_s^+(\R^n)$, since in this case  inequality  \fer{posi} is no more valid. 

It is remarkable that the functional $T_n(F)$  defines a measure of inequality for multivariate densities which satisfies most of the properties satisfied by its one-dimensional version $T(F)$.  

Proceeding as in the one-dimensional case, we can check that the upper bound in \fer{bound-n} is reached simply by evaluating the value of the multivariate index in correspondence to a multivariate random variable $\bf X$ taking only the two values $\aa= (a_1,a_2, \dots, a_n)$ and $\ab= (b_1,b_2,\dots, b_n)$ in $\R^n$ with probabilities $1-p$ and, respectively $p$, where $0<p<1$. As we will see in Section \ref{sec:Gini-n}, this example clarifies the advantages in measuring multidimensional heterogeneity by means of this new index, with respect to the use of existing generalizations of Gini index to the multidimensional setting \cite{AZ,Ba19}.  For this reason, we will give it into details. \\ The Fourier transform of the distribution $F$ of $\bf X$ is given by
\be\label{two-points}
\ff(\bxi) = (1-p)e^{-i\aa\bxi} + p e^{-i \ab\bxi}.
\ee
Consequently
\[
\nabla \ff(\bxi) = -i \left[ (1-p)\aa^T e^{-i\aa\bxi} + p\ab^T e^{-i \ab\bxi} \right] 
\]
and 
\[
\nabla \ff(\bxi = {\bf 0}) = -i \left[ (1-p)\aa^T  + p\ab^T  \right],
\]
so that
\[
\nabla \ff(\bxi = {\bf 0}) \ff(\bxi) - \nabla \ff(\bxi) = i\,p(1-p) (\ab^T -\aa^T)\left[e^{-i\aa\bxi} - e^{-i \ab\bxi}\right].
\]
Therefore
\begin{equations}\nonumber
T_n(F) =&\frac 1{2|\bm|} p(1-p) |\ab^T-\aa^T|  \sup_{\bxi \in \R^n} \left| e^{-i\aa\bxi} - e^{-i \ab\bxi}\right| = \\
&\frac 1{2|\bm|} p(1-p) |\ab^T-\aa^T|  \sup_{\bxi \in \R^n} \left| 1 - e^{-i (\ab-\aa)\bxi}\right| = 
 \frac 1{|\bm|} p(1-p) |\ab^T-\aa^T| .
\end{equations}
Hence, expanding the value of the mean $\bm$ we get the formula
\be\label{two}
T_n(F) = \frac{p(1-p) |\ab^T-\aa^T|}{|(1-p)\aa^T  + p\ab^T |}.
\ee
This expression has a structure which does not differ from the the one-dimensional formula computed  in \cite{To1}, that reads
\[
T(F) = T_1(F) = \frac{p(1-p) |b-a|}{(1-p)a  + p b}.
\]
In fact, when the mean is fixed, the value of the index does not depend on the positions of the two points $\aa$ and $\ab$, but only on their distance. \\
To show that formula \fer{two} can be used to reach the upper bound, let us now consider the case in which, for a given positive constant $\e\ll 1$, $p =\e$,  the point $\aa ={\bf 0}$,  while $\ab = \bm/\e$ is located far away, but leaving the mean value $\bm$ unchanged. In this case $T_n(F) = 1-\e$, a value which, as $\e \to 0$  converges to the upper bound expressed by the value $1$.

As its one-dimensional version, we can further show that the inequality index $T_n$ satisfies a number of properties \cite{GRT}.

Let $F,G\in \tilde P_s(\R^n)$ two probability measures with the same mean value, say $\bm$. Then, for any given $\t \in (0,1)$ it holds 
\be\label{Conv}
T_n (\t F + (1-\t)G)  \le 
 \t\, T_n(F) + (1-\t)T_n(G).
 \ee
 Inequality \fer{Conv} shows the convexity of the functional $T_n$ on the set of probability measures with the same mean.
 
Another interesting  property characterizing the inequality index $T_n$ is linked to its behavior when evaluated on convolutions. Let $\bf X$ and $\bf Y$  independent multivariate random variables with probability measures in $\tilde P_s(\R^n)$, and mean values $\bm_X$ (respectively $\bm_Y$).  Then if $\ff(\bxi)$ and $\gg(\bxi)$ denote the Fourier transforms of their respective probability measures, the Fourier  transform $\hh(\bxi)$ of the distribution measure of the sum ${\bf X}+ \bf Y$ is equal to the product $\ff(\bxi)\gg(\bxi)$.

Then, it can be shown that \cite{GRT} 
\be\label{convo}
T_n({\bf X}+ \bf Y) \le  \frac{|\bm_X|} {|\bm_X +\bm_Y|} T_n({\bf X}) + \frac{|\bm_Y|} {|\bm_X +\bm_Y|} T_n({\bf Y}).
\ee
It is remarkable that, at difference with the one-dimensional case, in \fer{convo} the sum of the coefficients in front of the inequality indices  $T_n({\bf X})$ and $T_n({\bf Y})$ is always greater that one. Nevertheless, one can extract from \fer{convo} some useful consequences.

In particular, if $\bf X$  and $\bf Y$ belong to $P_s^+(\R^n)$ and $\bf Y$ is a random variable that takes the value $\bm$ with positive components with probability $1$ (so that $\gg(\bxi) = e^{-i\bm\bxi}$ and $T({\bf Y}) = 0$), 
\be\label{tide}
T_n({\bf X}+ {\bf Y}) =  \frac{|\bm_X|} {|\bm_X +\bm_Y|} T_n({\bf X}) < T_n(\bf X),
\ee
since in this case the length of the sum of two vectors with positive components is bigger than the length of both. It is remarkable that the same result holds even if only one component of the $\bf Y$ variable is bigger than zero, while the others are not. The meaning of inequality \fer{tide} is clear. 
Since in this case $\bf X+Y$ is nothing but $\bf X +\bm$, which corresponds to adding  constant values $m_k$ to $X_k$, for $k=1, 2,\dots, n$, this property asserts that adding a positive constant value to one or more components  to each agent decreases inequality. 

Also, if the  independent random variables ${\bf X}_1$ and ${\bf X}_2$ are distributed with the same law of $\bf X$, so that their mean values are equal, thanks to the scale property 
\be\label{clt}
T_n\left(\frac{{\bf X}_1+{\bf X}_2}2\right) = T_n\left({\bf X}_1+{\bf X}_2\right) \le T_n (\bf X),
\ee
while the mean of $({\bf X}_1+{\bf X}_2)/2$ is equal to the mean of $\bf X$.

A third important consequence of inequality \fer{convo} is related to the situation in which the random variable ${\bf Y} ={\bf N}$ represents a noise (of  mean value $\bm>0$) that is present when measuring the inequality index of $\bf X$.  The classical choice is that the additive noise is represented by a Gaussian variable of mean $\bm$ and covariance matrix $\Sigma$. 

We have in this case
\be\label{new-gau}
T_n({\bf X}+ {\bf N}) \le \frac{|\bm_X|} {|\bm_X +\bm|} T_n({\bf X}) + \frac{|\bm|} {|\bm_X +\bm|} T_n({\bf N}).
\ee
Hence, a precise upper bound can be obtained once we know the explicit value of the inequality index $ T_n({\bf N})$. This leads to the interesting question relative to the (explicit) evaluation of the inequality index $T_n$ of a random multivariate Gaussian variable.

This evaluation has been done in \cite{GRT}. 
The Fourier transform of the distribution function $F$ of a multivariate Gaussian variable ${\bf N} = (N_1,N_2,\dots, N_n)$ in $\R^n$, $n >1$, is given by the expression
\be\label{gauss}
\ff(\bxi) = \exp\left\{-i \bm^T\bxi - \frac 12 \bxi^T\Sigma \, \bxi \right\}, 
\ee
where $\bm$ is the vector of the mean values $\langle N_k\rangle$ , and $\Sigma$ is the $n\times  n$ covariance matrix, with elements
\[
\sigma_{ij} = \langle(N_i -m_i)(N_j-m_j)\rangle.
\]
Then, for the multivariate Gaussian one obtains the expression
\be\label{general}
T_n(N) = \frac 1{2\sqrt e} \frac 1{|\bm|} \sqrt{\frac{\sum_{k=1}^n \lambda_k^2}{\sum_{k=1}^n \lambda_k}},
\ee
where  the $\lambda_k$'s are the positive eigenvalues of the covariance matrix of the multivariate Gaussian  distribution. \\
Going back to formula \fer{new-gau}, in presence of a additive noise  represented by a Gaussian variable of mean $\bm$ and covariance matrix $\Sigma$  one has the bound
\be\label{gau-expl}
T_n({\bf X}+ {\bf N}) \le \frac{|\bm_X|} {|\bm_X +\bm|} T_n({\bf X}) + \frac 1{2\sqrt e|\bm_X +\bm|}\sqrt{\frac{\sum_{k=1}^n \lambda_k^2}{\sum_{k=1}^n \lambda_k}} .
\ee
It is interesting to remark that formula \fer{gau-expl} continues to hold in presence of a centered Gaussian random noise of covariance matrix $\Sigma$,
and in this case
\be\label{gau-cent}
T_n({\bf X}+ {\bf N}) \le  T_n({\bf X}) + \frac1{2\sqrt e|\bm_X| }\sqrt{\frac{\sum_{k=1}^n \lambda_k^2}{\sum_{k=1}^n \lambda_k}} .
\ee

\section{About the Gini-type index for multivariate distributions}\label{sec:Gini-n}

 Sections \ref{sec:nd} has been devoted to the definition of a new multivariate inequality index, to its main properties, and to its evaluation in correspondence to a multivariate Gaussian distribution. This analysis takes a great advantage from the possibility to express the index in terms of a multidimensional Fourier transform. \\ It is therefore fair to ask whether the use of the Fourier transform can also bring advantage in the definition of a multivariate Gini index. As a matter of fact, the extension of the classical Gini index to measure inequality in multivariate distributions has shown numerous attempts, as certified by the references of the recent paper \cite{Ba19}, in which the author is motivated by  the objective of designing a multidimensional Gini index  of inequality, to quantify of standard of living, that would satisfy all of a number of reasonable properties. This in reason of the fact that, as noticed in the introduction of \cite{Ba19},  many existing multidimensional inequality indices of Gini type proposed by economists from time to time have remained elusive in this respect. \\
To better understand the difficulties that appear when trying to build a multidimensional generalization of the Gini index, following the line considered in this paper, we will build a multivariate version of the index obtained by resorting to its Fourier one-dimensional transform. Indeed, the Fourier expression of the one-dimensional Gini index considered in \cite{To1} appears ready to be extended to higher dimensions, still preserving scale invariance.

As shown in Section \ref{sec:nd}, for any probability measure $F\in P_s^+(\R)$ of mean value $m>0$,  Gini index has  a simple expression in Fourier transform, given by \fer{Gin-F}.\\
Considering that the value zero in \fer{Gin-F} is obtained when $\ff(\xi) = e^{-im\xi}$,   Gini index can be fruitfully rewritten as
\be\label{G-diff}
G(F) =  \frac 1{2\pi m} \left[ \int_\R \frac{|1 - e^{-im\xi} |^2}{\xi^2} \, d\xi - \int_\R \frac{|1 -\ff(\xi)|^2}{\xi^2} \, d\xi\right].
\ee
Taking into account that inequality measures are scale invariant, expression \fer{G-diff} can be easily extended to measure the inequality of a multivariate distribution $F\in P_s^+(\R^n)$, $n >1$   by setting
\be\label{G_n}
G_n(F) =  \frac{\mu_n}{|\bm|} \left[ \int_{\R^n} \frac{|1 - e^{-i\bm\bxi} |^2}{|\bxi|^{n+1}} \, d\bxi - \int_{\R^n} \frac{|1 -\ff(\bxi)|^2}{|\bxi|^{n+1}} \, d\bxi\right], 
\ee
where the constant $\mu_n$ is such that
\be\label{con}
\frac 1{\mu_n}= \frac 1{|\bm|} \int_{\R^n} \frac{|1 - e^{-i\bm\bxi} |^2}{|\bxi|^{n+1}} \, d\bxi .
\ee
Evaluating the integral on the right-hand side by resorting to a $n$-dimensional spherical coordinate system, one realizes that the value of the constant $\mu$ does not depend on the vector $\bm$, and 
\[
\mu_n =\Gamma\left(\frac{n-1}2 +1 \right)\sqrt{(2\pi)^{n-1}},
\]
where $\Gamma (\cdot)$ denotes as usual the Gamma function. Formula \fer{G_n} is valid for all values of $n \in \N$, including $n =1$, that consistently gives $\mu_1=1$. \\  Resorting to \fer{con}, we can express the multivariate Gini-type index \fer{G_n} in the (simpler) form
\be\label{G-n}
G_n(F) =  1-  \frac{\mu_n}{|\bm|} \int_{\R^n} \frac{|1 -\ff(\bxi)|^2}{|\bxi|^{n+1}} \, d\bxi.
\ee
The same idea can be applied to recover an expression for a multivariate Pietra index \cite{Pie, To1}. However, despite their eventual theoretical interest, does not seem that this type of expressions, if compared to the multivariate $T_n$ index considered in this paper, share good properties.

The problems which appear when passing from the one-dimensional version \fer{Gin-F} to its natural multivariate version \fer{G-n} can be easily understood by evaluating  the value of the Gini index $G_n$, $n >1$, in correspondence to the multivariate random variable $\bf X$ taking only two values,  introduced in Section \ref{sec:nd}. This variable is characterized by the Fourier transform \fer{two-points}, so that,  to compute the value of  Gini index, as expressed by formula \fer{G-n}, we need to evaluate the integral
\[
I_n(F)= \frac {\mu_n}{|\bm|} \int_{\R^n} \frac{|1 -  (1-p)e^{-i\aa\bxi} + p e^{-i \ab\bxi}|^2}{|\bxi|^{n+1}} \, d\bxi,
\]
where $|\bm| = |(1-p)\aa^T  + p\ab^T |$. It is immediate to show that the integral $I_n$ can be split into three terms, i.e.
\begin{equations}\label{I-n}
I_n(F) &= \frac{\mu_n}{|\bm|} \int_{\R^n} \frac{2(1-p) (1 - \cos \aa \bxi)}
{|\bxi|^{n+1}} \, d\bxi + \\
& \frac{\mu_n}{|\bm|} \int_{\R^n} \frac{2p (1 - \cos \ab \bxi)}{|\bxi|^{n+1}} \, d\bxi + \frac{\mu_n}{|\bm|} \int_{\R^n} \frac{2p(1-p) (1 - \cos (\ab-\aa) \bxi)}
{|\bxi|^{n+1}} \, d\bxi.
\end{equations}
The three integrals on the right-hand side of \fer{I-n} can be easily evaluated  by resorting to a $n$-dimensional spherical coordinate system to give
\[
I_n(F) = \frac{1}{|\bm|}\left[ (1-p)|\aa^T| + p|\ab^T| - p(1-p)|(\ab^T - \aa^T |\right],
\]
so that 
\begin{equations}\label{bad}
G_n(F) &= 1 -\frac{1}{|\bm|}\left[ (1-p)|\aa^T| + p|\ab^T| - p(1-p)|(\ab^T - \aa^T |\right]=\\
& \frac{|(1-p)\aa^T  + p\ab^T | - \left[ (1-p)|\aa^T| + p|\ab^T|\right] + p(1-p)|(\ab^T - \aa^T |}{|(1-p)\aa^T  + p\ab^T |} \le \\
& \frac{ p(1-p)|(\ab^T - \aa^T |}{|(1-p)\aa^T  + p\ab^T |}  = T_n(F).
\end{equations}
Hence, at difference with the $T_n$ multivariate index, the multivariate Gini index $G_n$ does not coincide, even in the case of a simple two-valued distribution, with the $n$-dimensional extension of the univariate index. In other words, while the unidimensional indices depend only on the modulus of the difference between the the two values assumed by the random variable, in the multivariate case only the $T_n$ index retains this property, while the multivariate Gini index $G_n$, apparently derived from a natural extension of the one-dimensional index by maintaining the scaling property, does not. \\
In fact the additional term appearing on the numerator of formula \fer{bad}, given by
\[
|(1-p)\aa^T  + p\ab^T | - \left[ (1-p)|\aa^T| + p|\ab^T|\right],
\]
even in presence of two vectors with positive components, in dimension $n >1$ is  dependent on the position of the points $\aa$ and $\ab$ on the space $\R^n$, and it is equal to zero if and only if the two vectors $\aa$ and $\ab$ are parallel. This unpleasant fact shows that even the passage to Fourier transform does not allow a simple extension of the Gini index to multivariate distributions. \\ On the contrary, the presence of heavy difficulties in defining a easy to treat inequality index able to measure multivariate distributions characterizes the $T_n$ index introduced in this paper as a good candidate for future applications.

\section{Conclusions}
The description of social phenomena in a multi-agent system by means of kinetic equations often leads to the identification of multidimensional universal steady profiles, equilibrium distributions of paramount importance that should resume at best the characteristics of the phenomenon under investigation, dependent  in general on several factors.
Among the various features considered to have a more precise measurement of the social characteristics of the steady profile, multivariate inequality indices represent a primary tool \cite{AA,Eu,Eu2,AZ}.\\
In this paper, we enlightened various properties of a new inequality index $T_n$, considered in \cite{GRT}, characterized in terms of the multidimensional Fourier transform, which appears to have a number of good properties in the general case of multivariate distributions.  The interest in applications of the index $T_n$, is amplified by the fact that the Fourier transform natural generalization of the one-dimensional Gini index to multivariate distribution does not lead to a definition which satisfies the basic properties required to inequality indices. 

\section*{Acknowledgements}

 This work has been written within the activities of GNFM (Gruppo Nazionale per la
Fisica Matematica) of INdAM (Istituto Nazionale di Alta Matematica), Italy. The research was partially
supported by the Italian Ministry of Education, University and Research (MIUR) through the “Dipartimenti
di Eccellenza” Programme (2018-2022) -- Department of Mathematics “F. Casorati”, University of Pavia.


\end{document}